      Text body of the manuscript

\documentstyle[12pt]{article}

\newtheorem{prop}{Proposition}

\newtheorem{thm}{Theorem}

\newcommand {\beq}{\begin{equation}}
\newcommand {\eeq}{\end{equation}}
\newcommand {\beqa}{\begin{eqnarray}}
\newcommand {\eeqa}{\end{eqnarray}}         
\newcommand {\beqs}{\begin{eqnarray*}}
\newcommand {\eeqs}{\end{eqnarray*}}
\newcommand {\bds}{\begin{displaymath}}
\newcommand {\eds}{\end{displaymath}}

\newcommand {\n}{\nonumber \\}
\newcommand {\nn}{\nonumber}
\newcommand{\no}{\noindent} 

\newcommand {\bebb}{\begin{thebibliography}}
\newcommand {\eebb}{\end{thebibliography}}      
\newcommand {\bbit}{\bibitem}
\newcommand{\slt}{sl_2}

\newcommand{\dyslt}{{{\cal D}Y(\slt)}}

\def\hb2{\frac{\hbar}{2}}

\def\al{\alpha}
\def\bt{\beta}

\def\gm{\gamma}
\def\dl{\delta}

\def\psh{\partial}

\def\bop{\bigoplus}

\def\rtarr{\rightarrow}

\def\ket#1{|#1\rangle}

\begin{document}

\baselineskip = 15 pt


\title{
\LARGE\bf Free Boson Representation of $DY_{\hbar}(sl_2)_k$
and the Deformation of the Feigin-Fuchs}
\vspace{1cm}
\author{
{\normalsize\bf
X.M. Ding $^{a,b}${\thanks {E-mail:xmding@itp.ac.cn}}, 
L. Zhao $^c${\thanks {E-mail :lzhao@nwu.edu.cn}}
}\\
\normalsize $^a$ CCAST, P.O. Box 8730, Beijing,100080, China\\
\normalsize $^b$ Institute of Theoretical Physics of Acedimia Sinica,
P.O.Box 2735, 100080, China\\
\normalsize  $^c$ Institute of Modern Physics, Northwest University, Xian, 
710069, China\\
}

\date{}

\maketitle

\vspace{2cm}

\begin{abstract}

A realization of the Yangian double with center $DY_\hbar(sl_2)_k$ of 
level $k(\not=0,-2)$ in terms of free
boson fields is constructed. The screening currents are also presented,   
which commute with $DY_\hbar(sl_2)$ modulo total difference. In the 
$\hbar\!\rightarrow\! 0$ limit, the currents of Yangian double 
$DY_\hbar(sl_2)_k$ becomes the Feigin-Fuchs realization of affine 
Lie $sl(2)_k$, while the screening currents of Yangian double 
$DY_\hbar(sl_2)_k$ becomes the screening currents of the affine Lie algebra 
$sl(2)_k$.   

\end{abstract}


\setcounter{section}{0}
\setcounter{equation}{0}
\section{Introduction}

Yangians were proposed by Drinfeld as generalizations
of classical loop algebra with nontrivial Hopf algebra structures
\cite{D1,D2,Dr:new}. From the physics point view, they are the symmetric  
algebra of the non-local currents of massive field theories \cite{BL}. 
Following the Faddeev-Reshetikhin-Takhtajan (FRT) 
formalism\cite{FRT}, as an associative algebras, Yangian can be defined 
through the Yang-Baxter relation (i.e. $RLL$-relations) with the structure 
constants determined by the rational solutions of the quantum Yang-Baxter
equation (QYBE). However, this is just one part of the story, 
for physical applications, such as description the dynamical symmetry of the 
massive integrable quantum filed theories, the dual modules and the infinity 
dimensional representations are required\cite{BL,LS,S2}, in another words, the 
Yangian double with center is emerged naturally as the dynamical symmetry of the 
integrable massive field theories. In fact, Yangian double 
with center were proved to have important applications in certain physical 
problems, especially in describing the dynamical symmetries of the 
perturbative integrable massive quantum field theories, calculating 
the correlation functions and form factors of some two-dimensional
exactly solvable lattice statistical model and (1+1)-dimensional
completely integrable quantum field theories \cite{BL,LS,KK,S2}. 

From the $RLL$ viewpoint, if one regards the 
Reshetikhin-Semenov-Tian-Shansky realization (RS)\cite{RS} as the affine  
extension of Faddeev-Reshetikhin-Takhtajan formalism. Yangian double with center 
(the central extension for the quantum double of Yangian) \cite{K,KL,KT} are 
affine extensions of Yangian. Contrary to the quantum affine (corresponding to  
trigonometric solution of QYBE) case, the Yangian double with center is only 
recently obtained \cite{IK,K,KL}.   

The free fields realization is a common used method in the quantum field theory. 
The free boson representations of $U_q(\widehat{sl_2})$ with an arbitrary
level have been obtained in Refs.\cite{Srs,M1,M2}.
For the Yangian double with center, the free field representation of 
$DY_\hbar(sl_2)$ with level $k(\neq 0 -2)$ was constructed in \cite{konno}, 
the level-$1$ and level-$k$ representation of  
$DY_{\hbar}(sl_N)$ are obtained in \cite{Ioh} and 
\cite{HZD}, respectively. The level-$k$ free field representation of 
$DY_{\hbar}(gl_N)$ is also given in \cite{DHHZ}. However, all of these are the 
deformation of the Wakimoto modules, there is another kind of 
module-Feigin-Fucks 
module in the classical \cite{Nem} and the quantum affine case \cite{BV,DW}. In 
the case of Yangian double with center this kind of module is not given yet.  

In this manuscript, we consider a representation of the level-$k\ (\not=0,-2)$ 
Drinfeld currents in terms of the three free bosonic fields $\bt, \al$ and
$\gm$. In the classical limit $\hbar \rtarr 0$, this module becomes the 
Feigin-Fuchs representation of the affine Lie algebra $sl(2)_k$. So this module 
can be regarded as a deformation of the so-called
Feigin-Fuchs construction of the affine Lie algebras\cite{FF,Nem}.
 
The manuscript is arranged as follows. Section $2$, at first, we briefly review 
the Drinfeld new realization of the Yangian $Y(sl_2)$, then we give the 
defining relations of Drinfeld generators of the Yangian double with center 
$DY_{\hbar}(gl_2)$. In section $3$, the three free boson fields are introduced; 
and a free fields realization of $DY_{\hbar}(sl_2)$ are given. The free fields 
realization of the screening currents is obtained in section $4$. 

\setcounter{section}{1}
\setcounter{equation}{0}
\section{Yangian and Yangian double with center}

Yangian is a Hopf algebra, there are three realizations of Yangian $Y(g)$: 
Drinfeld realization \cite{D1,D2}, Drinfeld new realization\cite{Dr:new}, 
and $R$-matrix realization or FRT realization \cite{KR,FRT}. 
The isomorphism between 
the Drinfeld new realization and the $R$-matrix (or FRT) realization can be 
established by using the Ding and Frenkel correspondence \cite{DF}. Here, we 
only give the definition of the Drinfeld new realization of Yangian 
$Y(\slt)$, for the Yangian double ${{\cal D}Y(\slt)}$
is defined in terms of quantum double of the Drinfeld new realization, 
while the Yangian double with center is the central extension of the 
Yangian double. There is the following theorem:

\begin{thm}\cite{Dr:new} Yangian $Y(\slt)$ is a Hopf algebra generated by the 
symbols $e_m, f_m, h_m\ m\geq 0$ satisfying the relations

\beqa
& &[h_m, h_n]=0, \qquad [e_m, f_n]=h_{m+n} ,\n
& &[h_0, e_m]=2e_m, \qquad [h_0, f_m]=-2f_m ,\n
& &[h_{m+1}, e_n]-[h_m, e_{n+1}]=\hbar\{h_m,e_n\}, \\
& &[h_{m+1}, f_n]-[h_m, f_{n+1}]=-\hbar\{h_m,f_n\}, \n
& &[e_{m+1}, e_n]-[e_m, e_{n+1}]=\hbar\{e_m,e_n\}, \n
& &[f_{m+1}, f_n]-[f_m, f_{n+1}]=-\hbar\{f_m,f_n\}.\nn
\eeqa

\no Here $\hbar$  is a formal variable and $\{a,b\}=ab+ba$. 
\end{thm}

\no The central extension of the Yangian double 
\cite{IK,KT} is a Hopf algebra over $C [[\hbar]]$ generated by 
$\{e_m, ~f_m, ~h_m, ~c, ~~d|~m\in Z \}$. The generating functions are 

\beqa 
& & e^{\pm}(u)=\pm\sum_{m\geq 0\atop m<0}e_m u^{-m-1},\qquad
 f^{\pm}(u)=\pm\sum_{m\geq 0\atop m<0}f_m u^{-m-1},\qquad
\n
& &h^{\pm}(u)=1\pm\hbar\sum_{m\geq 0\atop m<0}h_m u^{-m-1}, 
\label{modes}
\eeqa
\no and 
\beqs
&& e(u)=e^+(u)-e^-(u),\qquad f(u)=f^+(u)-f^-(u),\\
&& \delta(u-v) = \sum_{n+m=-1}u^nv^m.\n
\eeqs
\no satisfying the following relations.

\beqa
& &[d, \chi(u)]=\frac{d}{du}\chi(u),\qquad 
\chi(u)=e(u),~~f(u),~~h^{\pm}(u)\n
& & e(u)e(v)=\frac{u-v+\hbar}{u-v-\hbar}e(v)e(u),\n
& & f(u)f(v)=\frac{u-v-\hbar}{u-v+\hbar}f(v)f(u),\n
& &h^{\pm}(u)e(v)=\frac{u-v+\hbar}{u-v-\hbar}e(v)h^{\pm}(u),\n
& & h^+(u)f(v)=\frac{u-v-(1+c)\hbar}{u-v+(1-c)\hbar}f(v)h^+(u),
\label{define}\\
& & h^-(u)f(v)=\frac{u-v-\hbar}{u-v+\hbar}f(v)h^-(u),\n
& & [ h^{\pm}(u), h^{\pm}(v) ]=0,\n
& & h^+(u)h^-(v)=\frac{u-v+\hbar}{u-v-\hbar}
        \frac{u-v-(1+c)\hbar}{u-v+(1-c)\hbar}h^-(v)h^+(u),\n
& & [e(u), f(v)]=\frac{1}{\hbar}
(\delta(u-(v+\hbar c))h^+(u)-\delta(u-v)h^-(v)),
\nn
\eeqa

\no and the coproduct can be found\cite{IK,KL} 

\setcounter{section}{2}
\setcounter{equation}{0}

\section{Free bosons}
For the following usage, we first introduce a Heisenberg algebra ${\cal H}$ 
generated by 
${\al}_n$, $\bt _n$, $\gm _n$, $n\in Z$ and $P_{\al}$, 
~$P_{\bt}$, $\ P_{\gm}$, ~$Q_{\al}$, ~$Q_{\bt}$, ~$Q_{\gm}$ 
satisfying the commutation relations

\beqa
&& [\al_m, \al_n ]=\frac{k+2}{2}m\delta_{m+n,0},\qquad 
 [ P_{\al}, Q_{\al}]=\frac{k+2}{2}, \n
&& [\bt_m, \bt_n ]=-\frac{k}{2}m\delta_{m+n,0},\qquad  
[ P_{\bt}, Q_{\bt}]=-\frac{k}{2}, \label{heisenberg}\\
&& [\gm_m, \gm_n ]=\frac{k}{2}m\delta_{m+n,0},\qquad  
[ P_{\gm}, Q_{\gm}]=\frac{k}{2},\nn
\eeqa

\no with $k\in C(\not=0,-2)$. The other commutators vanish identical. 
Denote the vacuum states with $\al, \bt, \gm$-charges
$l, s, t$ as $\ket{l;s,t}$ : 

\beqs
& &\ket{l;s,t}=e^{\frac{l}{k+2}Q_{{\al}}-\frac{s}{k}Q_{\bt}
+\frac{t}{k}Q_{\gm}}\ket{0},\\
& & X_n\ket{0}=0,\quad n{>0}, \quad P_{X}\ket{0}=0,
\eeqs

\no for $X=\al,\bt,\gm$, and as ${\cal F}_{l,s,t}$ the Fock space 
constructed  on $\ket{l;s,t}$:

\beq
{\cal F}_{l,s,t}=\Bigl\{\prod \al_m \prod \bt_n\prod \gm_r\ket{l;s,t}\Bigr\}
\eeq

\no with $m, n, r\in  Z_{<0}$. It is convenient to introduce generating 
functions $X( u; A,B)$ ($X=\al, \bt, \gm$) as

\beqa
X(u; A,B)=&&\sum_{n>0}\frac{X_n}{ n}(u+A\hbar)^{n}
              -\sum_{n>0}\frac{X_n}{n}(u+B\hbar)^{-n} \nonumber \\ 
&&~~~ +  \log(u+B\hbar)P_{{X}}+Q_X.
\eeqa

\no and use the notation $X(u; A,A)=X(u;A)$ for concise. The normal ordered 
product $:\ :$ is defined as moving the positive frequencies to the right 
of the negative ones. While the difference operator ${}_{\al}{\psh}_u f$ is 
defined as:

\bds
_{\al}\psh _{u}f(u)=\frac{f(u+\al\hbar)-f(u)}{\hbar}.
\eds

\no Then we have the following results:

\begin{prop}\label{}: The currents $e(u), f(u)$ and $h^{\pm}(u)$ of the 
Heisenberg algebra ${\cal H}$ acting on ${\cal F}_{l,s,t}$ can be defined by 

\beqa
 e(u)&=&-\frac{1}{\hbar}: \Bigl[ \exp\Bigl\{ \sum _{n>0}\frac{1}{n}
(\al _n +\bt _n)\bigl[ (u-(k+1)\hbar)^{-n}-(u-(k+2)\hbar)^{-n}\bigr]\Bigr\} \n
& &~~~~~\times
\bigl( \frac{u-(k+2)\hbar}{u-(k+1)\hbar} \Bigr)^{P_{\al}+P_{\bt}} \n
& &~~-
\exp \Bigl\{ \sum_{n>0}\frac{1}{n}
(\al _{-n} +\bt _{-n}) [(u-(k+1)\hbar)^{n}-(u-(k+2)\hbar)^{n}]\Bigr\}\Bigr]\n
& &~~\times\exp\Bigl\{\frac{2}{k}(\bt +\gm)(u; -(k+1),-(k+2)) \Bigr\}:,
\label{dre}\\
f(u)&=&\frac{1}{\hbar}
:\Bigl[ \exp\Bigl\{\sum_{n>0}\frac{1}{n}
(\al _n +\bt _n)[(u-2\hbar)^{-n}-(u-\hbar)^{-n}]\Bigr\}
\bigl(\frac {u-\hbar}{u-2\hbar}\Bigr)^{P_{\al}+P_{\bt}}\n
& &~~~~~\times\exp\Bigl\{\sum_{n>0}\frac{1}{n}
\gm _n[(u-2\hbar)^{-n}-u^{-n}]\Bigr\}
\bigl(\frac{u}{(u-2\hbar)}\Bigr)^{P_{\gm}}\n
& &~~\times\exp\Bigl\{-\frac{2}{k}(\bt +\gm)(u; -1,-2) \Bigr\}\n
& &~~-\exp\Bigl\{\sum_{n>0}\frac{1}{n}
(\al _{-n} +\bt _{-n})[(u-(k+2)\hbar)^{-n}-(u-(k+3)\hbar)^{-n}]\Bigr\}\n
& &~~\times\exp\Bigl\{\sum_{n>0}
\frac{2\al _n}{(k+2)n}[(u-(k+3)\hbar)^{-n}-(u-\hbar)^{-n}]\Bigr\}\n
& &~~~\exp\Bigl\{-\frac{2}{k}(\bt +\gm)(u; -1,-(k+2)) \Bigr\}\Bigr]:,
\label{drf}\\
h^+(u)&=&
\exp\Bigl\{  \sum_{n>0}\frac{\gm_n}{n}[(u-(k+2)\hbar)^{-n}-
(u-k\hbar)^{-n}]\Bigr\} 
\Bigl(\frac{u-k\hbar}{u-(k+2)\hbar}\Bigr)^{P_{\gm}}
\label{dhp}\\
h^-(u)&=&\exp\Bigl\{ \sum_{n>0}\frac{2}{nk}(\bt {-n}+\gm _{-n})
[(u-(k+1)\hbar)^{n}-
(u-\hbar)^{n}]\Bigr\}\n
& & \qquad\times \exp\Bigl\{- \sum_{n>0}\frac{1}{n}
(\al_{-n}+\bt _{-n})[(u-(k+3)\hbar)^{n}-
(u-(k+1)\hbar)^{n}]\Bigr\}\n
& &~~~~~ \times \exp\Bigl\{\sum_{n>0}\frac{2}{(k+2)n}
\al_{-n}
[(u-(k+3)\hbar)^{n}-(u-1\hbar)^{n}]\Bigr\}
\label{dhn}
\eeqa

\no and $d$ operator by

\beq
d= \frac{2}{k+2}d_{\al}-\frac{2}{k}d_{\bt}+\frac{2}{k}d_{\gm},
\label{deriv}
\eeq

\no in which 

\beqs
&& d_{\al}=\al_{-1}P_{\al}+
\sum_{n\in Z_{>0}}\al_{-n-1}\al_n,\\
&& d_{\bt}= \bt_{-1}P_{\bt}+ \sum_{n\in Z_{>0}} \bt_{-n-1}\bt_{n},\\
&&d_{\gm}=\gm_{-1}P_{\gm}+
\sum_{n\in Z_{>0}}\gm_{-n-1}\gm_{n}.
\eeqs
\end{prop}
\no And the following results are straightforward:



\begin{thm} {\it By analytic continuation, the Heisenberg algebra ${\cal H}$ of 
the 
currents $e(u), f(u)$ $h^{\pm}(u)$ and the operator $d$ are homomorphism    
(\ref{define}) with $c=k$ on ${\cal F}_{l,s,t}.$}
\end{thm}

\no So we can denote the Heisenberg algebra ${\cal H}$ generated by the 
currents $e(u), f(u), h^{\pm}(u)$ and $d$ on ${\cal F}_{l,s,t}$ as $\dyslt$.
 
In the classical limit $\hbar\to 0$, the currents $e(u), f(u),
h^{\pm}(u)$ tend to the Feigin-Fucks construction of the level-$k$ 
currents for the affine Lie algebra $\widehat{\slt}$:

\beqs
&& e(u)\to \Bigl(\sqrt{\frac{k+2}{2}}\psh \phi _1(u)
+\sqrt{ \frac{k}{2}}i\psh \phi _2(u)
\exp \Bigl\{\sqrt{\frac{2}{k}}(i\phi _2(u)-\phi _0(u) \Bigr\},\\
&& f(u)\to \Bigl(\sqrt{\frac{k+2}{2}}\psh \phi _1(u)
-\sqrt{ \frac{k}{2}}i\psh \phi _2(u)
\exp \Bigl\{-\sqrt{\frac{2}{k}}(i\phi _2(u)-\phi _0(u) \Bigr\},\\
&& \frac{1}{\hbar}\Bigl( h^+(u)-h^-(u)\Bigr) \to  
-\sqrt{\frac{k}{2}}\psh \phi _0(u),
\eeqs

\no where $\phi _i(u)\phi _j(v)=\dl _{i,j}\ln(u-v)$, and the operator $d$ is  
$L_{-1}$ of Virasoro algebra. 

It is easy to show

\bds
[\chi(u), P_{\bt}+P_{\gm}]=0,\quad \chi(u)=e(u),~~f(u),~~h^{\pm}(u).
\eds 

\no so in the following, we restrict the Fock space ${\cal F}_{l,s,t}$ to the 
$s=-t$ sector without loss of generality, on this sector the currents $e(u), 
f(u)$  and $h^{\pm}(u)$ are single valued so that the expansion such as 
(\ref{modes}) makes sense.

\setcounter{section}{3}
\setcounter{equation}{0}
\section{Screening currents}

In this section, we construct two screening currents of the algebra 
$\dyslt$. In the classical limit $\hbar\to 0$, they become the screening  
currents of the affine Lie algebra $sl(2)_k$. Let us next consider the following 
operators.

\bds
\xi(u)=S^+(u)^{-1}, 
\qquad S^+(u)=:\exp\{(\al +\bt) (u; -(k+2))\}: 
\eds

We have 

\beq
\xi(u)S^+(v)=-S^+(v)\xi(u)\sim\frac{1}{u-v}.
\label{xis}
\eeq

\no Here $\sim$ means that the relation is equivalent up to modulo regular 
terms.
The fields $\xi(u)$ and $S^+(u)$ are single valued on 
${\cal F}_{l,s,s}\ s\in Z$. From (\ref{xis}),
the zero-modes $\xi_0=\oint\frac{du}{2\pi i u}\xi(u)$ and 
$Q^{+}=\oint\frac{du}{2\pi i }S^+(u)$ anticommute
$ \{\xi_0, Q^+\}=0$. Note also $\xi^{2} _{0} =0 =Q_{+}^{2}$.
In addition,  the following equations hold in the sense of analytic 
continuation.

\begin{prop}

\beqs
&& h^{\pm}(u)S^+(v)=S^+(v)h^{\pm}(u)\sim 0,\\
&& f(u)S^+(v)=-S^+(v)f(u)\sim 0,\\
&& e(u)S^+(v)=-S^+(v)e(u) \\
&&\qquad\qquad\sim _{1} \psh _{v}
\Bigl(\frac{1}{u-v+\hbar}
\exp\{\frac{2}{k}(\bt +\gm) 
(u;-(k+2),-(k+3))\}\n
&&\qquad\qquad~~~~\times\exp\{(\al +\bt ) 
(u;-(k+2),-(k+3))\}~ \Bigr).
\eeqs
\end{prop}

\no Therefore the zero-mode $\eta_0$ commutes
with the action of  $\dyslt$. So the current $S^+(u)$ is the screening 
operator of $\dyslt$. we then restrict the Fock space ${\cal F}_{l,s,s}$ to the 
kernel of $\_0$, and hence arrive at the deformation of the Feigin-Fuchs modules

\beq
{\cal F}_l=\bop_{s\in Z}{\rm Ker}(Q^+: {\cal
F}_{l,s,s}\to 
{\cal F}_{l,s,s+1}).
\eeq

\no For the level$-k$ $U_q(\widehat{sl_2})$, the $q-$deformation of the
Feigin-Fuchs modules were obtained in \cite{BV,DW}.
 
Next we consider another screening operator defined by
\beqa
S(u)_{[L,M]}=\frac{1}{\hbar}:\Bigl[\exp\Bigl\{\sum_{n>0}
\frac{1}{n}(\al _n +\bt _n)\bigl[(u-\hbar})^{-n}-u^{-n}\bigr]\Bigr\}
\Bigl(\frac{u}{u-\hbar}\Bigr)^{P_{\al + P_{\bt}}
\n
-\exp\Bigl\{\sum_{n>0}
\frac{1}{n}(\al _{-n} +\bt _{-n)}\bigl[(u-\hbar)^{n}-u^{n}\bigr]\Bigr\}\Bigr]
\n
\times \exp\Bigl\{ -\sum_{n>0}\frac{2}{(k+2)n} \al _{-n} (u-\hbar)^{n}\Bigr\}
e^{ -\frac{2}{k+2} Q_{\al} }
\n
\times (u-2\hbar)^{\frac{2}{k}(P_{\bt} +P_{\gm})}
\prod _{l=1}^L \Bigl(\frac{u-(2+(k+2)l)\hbar}{u-(k+2)l\hbar}
\Bigr)^{\frac{2}{k}(P_{\bt} +P_{\gm})}
\n
\times \exp\Bigl\{\sum_{l=1}^L\sum_{n>0}
\frac{2}{kn}(\bt _n +\gm _n)\bigl[(u-(k+2)l\hbar)^{-n}
\n
-(u-(2+(k+2)l)\hbar)^{-n}\bigr]\Bigr\}
\n
\times \prod _{m=0}^M \Bigl(\frac{u-(2+(k+2)m)\hbar}{u-(k+2)m\hbar}
\Bigr)^{P_{\al}+P_{\bt} +P_{\gm}}
\n
\times \exp\Bigl\{\sum_{m=0}^M\sum_{n>0}
\frac{1}{n}(\al _n +\bt _n +\gm _n)\bigl[(u-(k+2)m\hbar)^{-n}
\n
-(u-(2+(k+2)m)\hbar)^{-n}\bigr]\Bigr\}:
\eeqa
\no with $L,M\in Z_{>0}$.

\no The following results are based on the direct calculation:

\begin{prop}

\beqs
h^{+}(u)S(v)_{[L,~M]}=S(v)_{[L,~M]}h^{+}(u)\sim 0,\\ 
e(u)S(v)_{[L,~M]}=S(v)_{[L,~M]}e(u)\sim 0,\\
S(u)_{[L,~M]}S^+(v)=S^+(v)S(u)_{[L,~M]}\\
\sim _{1}\psh _{v}
\Bigl(\frac{1}{u-v+(k+2)\hbar}
\exp\Bigl\{(\al  +\bt )(v;-(k+3),-(k+2))\Bigr\}
\\
\times \exp\Bigl\{ -\sum_{n>0}\frac{2}{(k+2)n} \al _{-n} (v)^{n}\Bigr\}
e^{ -\frac{2}{k+2} Q_{\al} }
\\
\times (v-\hbar)^{\frac{2}{k}(P_{\bt} +P_{\gm})}
\prod _{l=1}^L \Bigl(\frac{u-(2+(k+2)l)\hbar}{v-(-1+(k+2)l)\hbar}
\Bigr)^{\frac{2}{k}(P_{\bt} +P_{\gm})}
\\
\times \exp\Bigl\{\sum_{l=1}^L\sum_{n>0}
\frac{2}{kn}(\bt _n +\gm _n)\bigl[(v-(-1+(k+2)l)\hbar)^{-n}
\\
-(v-(1+(k+2)l)\hbar)^{-n}\bigr]\Bigr\}
\\
\times \prod _{m=0}^M \Bigl(\frac{v-(1+(k+2)m)\hbar}{v-(-1(k+2)m)\hbar}
\Bigr)^{P_{\al}+P_{\bt} +P_{\gm}}
\\
\times \exp\Bigl\{\sum_{m=0}^M\sum_{n>0}
\frac{1}{n}(\al _n +\bt _n +\gm _n)\bigl[(v-(-1(k+2)m)\hbar)^{-n}
\\
-(v-(1+(k+2)m)\hbar)^{-n}\bigr]\Bigr\}:
\\
\bigl[ S(u)_{[L,~M]},P _{\bt}+P_{\gm} \bigr]=0.
\eeqs

\no{\it In addition, in the limit $L,~M\to \infty$,}

\beqs
h^{-}(u)S^{}(v)=S^{}(v)h^{-}(u)\sim 0,\\
f(u)S(v)=S(v)f(u) \\
\qquad \sim _{k+2} \psh _{v}
\Bigl(\frac{1}{u-v}
\exp\Bigl\{ -\sum_{n>0}\frac{2}{(k+2)n} \al _{-n} (v-\hbar)^{n}\Bigr\}
e^{ -\frac{2}{k+2} Q_{\al} }
\n
\times (v-2\hbar)^{\frac{2}{k}(P_{\bt} +P_{\gm})}
\prod _{l=1}^L \Bigl(\frac{v-(2+(k+2)l)\hbar}{v-(k+2)l\hbar}
\Bigr)^{\frac{2}{k}(P_{\bt} +P_{\gm})}
\n
\times \exp\Bigl\{\sum_{l=1}^L\sum_{n>0}
\frac{2}{kn}(\bt _n +\gm _n)\bigl[(v-(k+2)l\hbar)^{-n}
\n
-(v-(2+(k+2)l)\hbar)^{-n}\bigr]\Bigr\}
\n
\times \prod _{m=0}^M \Bigl(\frac{v-(2+(k+2)m)\hbar}{v-(k+2)m\hbar}
\Bigr)^{P_{\al}+P_{\bt} +P_{\gm}}
\n
\times \exp\Bigl\{\sum_{m=0}^M\sum_{n>0}
\frac{1}{n}(\al _n +\bt _n +\gm _n)\bigl[(v-(k+2)m\hbar)^{-n}
\n
-(v-(2+(k+2)m)\hbar)^{-n}\bigr]\Bigr\}\Bigr)
\eeqs

\no {\it where $S(u)=\lim_{L,~M\to \infty}S(u)_{[L,M]}$.}
\end{prop}

\no Therefore the screening charge $Q=\oint\frac{du}{2\pi i}S(u)_{[L,~M]}$ 
commutes with all the currents in $\dyslt$ and $Q_+$
in the limit $L, ~M \to\infty$.
The charge $S$ yields a linear map $S: {\cal F}_{l}\to {\cal F}_{l-2}$.

We have constructed a free field representation of the level-$k$
Drinfeld currents for the Yangian double $\dyslt$ and screening operators.
As a result, we have obtained a deformation of the level-$k$ Feigin-Fuchs 
modules.

A possible application of the results is a calculation of
correlation functions in massive integrable quantum field theory such
as higher spin $SU(2)$ invariant Thirring model and in higher spin XXX
spin chains. For this purpose, one has to make a precise identification
of the space of states with the Feigin-Fuchs modules. We hope to discuss  
these problems in future publication. 
 
\vskip 1cm

\no {\bf Acknowledgments:} 
One of the authors (Ding) would like to thanks Prof. B. Y. Hou, prof K. Wu and 
Prof. Z. Y. Zhu for fruitful discussion, and he is supported in part by 
the "China postdoctoral Science Foundation".


\begin{thebibliography}{99}

 

\bibitem{BL}
Bernard, D., LeClair, A.: {\it Nucl. Phys.} {\bf B399} (1993), 709.

\bibitem{BV}
Bougourzi, A.H., Venet, L. :{\it Lett. Math. Phys. {\bf 36}} (1996), 101Nem.

\bibitem{DF}
Ding, J., and Frenkel, I.B.:
{\it Commun. Math. Phys.} {\bf 156} (1993), 277.


\bibitem{DHHZ}
Ding, X. M., Hou, B. Y. Hou, B. Yuan, Zhao, L.: Free Boson Representation of 
$DY_{\hbar}(gl_N)_k$ {\it Preprint{it hep-th/9709016}}.
 
\bibitem{DW}
Ding, X.M., Wang, P.: {\it Mod. Phys. Lett.{\bf A11}} (1996), 921.

\bibitem{D1}
Drinfeld, V.G.: {\it Soviet Math. Dokl.} {\bf 283} (1985), 1060.

\bibitem{D2}
Drinfeld, V.G.: In  {\it
Proceedings of the International Congress of Mathematicians},
p798, Berkeley, (1987).

\bibitem{Dr:new}
Drinfeld, V.G.: {\it Sov. Math. Dokl.} {\bf 36} (1988), 212.

\bibitem{FRT}
Faddeev, L.D., Reshetikhin, N.Yu., Takhtajan, L.A.:
{\it Advanced Series in Mathematical Physics}, {\bf Vol.10}, Singapore,
World Scientific (1989), p299.

\bibitem{FF}
Feigin, B.L., Frenkel, E.: {\it Russ. Math. Surv. {\bf 43}} (1989), 221.




 

\bbit{HZD}
Hou, B.Y., Zhao, L., X.M. Ding, :{\it Preprint{it q-alg/9701025}}.

\bibitem{Ioh}
Iohara, K. :{\it J. Phys.A.: Math. Gen.{\bf 29}},(1997),4593. 

\bibitem{IK}
Iohara, K., Konno, M.:
{\it Lett. Math. Phys. {\bf 37}} (1996), 319.


\bibitem{K}
Khoroshkin, S.: Central Extension of the Yangian Double. In
Collection SMF, Colloque ``Septi\`emes Rencontres du Contact Franco-Belge
en Alg\`ebre", June 1995, Reins; {\it Preprint {\tt q-alg/9602031}}.

\bibitem{KL}
Khoroshkin, S., Lebedev, D.: Intertwining operators for the
central extension of the Yangian double, {\it Preprint {\tt q-alg/9602030}}
(1996).

\bibitem{KLP}
Khoroshkin, S., Lebedev, D., Pakuliak, S.: Elliptic algebra ${\cal A} _{p,q}$ 
in the scaling limit, Preprint {\tt q-alg/9702002}.

\bibitem{KT}
Khoroshkin, S., Tolstoy, V.: {\it Lett. Math. Phys. {\bf 36}} (1996), 373.

\bibitem{KR}

Kirillov, A. N, Reshetikhin, N. Yu.: {\it Lett. Math. Phys. {\bf 12}} (1986), 
199.

\bibitem{KK}
Kojima, T., Korepin, V.E.: Determinant representations for
dynamical correlation functions of the quantum nonlinear
Schrodinger equation, {\it Preprint {\tt RIMS-1115}} (1996).

\bibitem{konno}
Konno, H.: Free field representation of level-$k$ Yangian double
$DY(sl_2)_k$ and deformation of Wakimoto Modules, {\it Lett. Math. Phys.  
{\bf 40}} (1997), 321.

{\it Preprint {\tt YITP-96-10}} (1996).

\bibitem{LS}
LeClair, A., Smirnov, F.:
{\it Int. J. Mod. Phys. {\bf A7}} (1992), 2997.

\bibitem{M1}
Matsuo, A.: {\it Phys. Lett. {\bf B308}} (1993), 260.

\bibitem{M2}
Matsuo, A.: {\it Commun. Math. Phys. {\bf 160}} (1994), 33.

\bibitem{Nem}
Nemeschansky, D.:{\it Phys. Lett. {\bf B242}} (1989), 121.

\bibitem{RS}
Reshetikhin, N. Yu., Semenov-Tyan-Shansky, M. A.:
{\it Let. Math. Phys. {\bf 19}} (1990), 133.

\bibitem{Srs}
Shiraishi, J.: {\it Phys. Lett. {\bf A171}} (1992), 243.

\bibitem{S2}
Smirnov, F.A.: {\it Int. J. Mod. Phys. {\bf A7 suppl. 1B}} (1992), 813, 839.

\bibitem{wakimoto}
Wakimoto, M.: {\it Commun. Math. Phys. {\bf 104}} (1986), 605.

\eebb

\end{document}